\documentclass[prc,aps,preprint,showpacs,nofootinbib,showkeys,eqsecnum]{revtex4}            
\usepackage{epsfig}
\newcommand{\noi}{\noindent}

\def\Journal#1#2#3#4{{#1} {\bf #2}, #3 (#4)}

\def\PRL{ Phys. Rev. Lett.\,\,}
\def\PRC{{ Phys. Rev.} \bf C}

\def\AJ{ Astrophys. J.\,\,}

\def\CPC{Comp. Phys. Comm.\,\,}
\def\JPBAMOP{J. Phys. B: At. Mol. Opt. Phys.\,\,}

\begin{document}
%
%
\title{
{\large{\bf Influence of protons on the capture of electrons by
the nuclei of $^{7}Be$ in the Sun}}}
\author{V. B. Belyaev}
\affiliation{Bogolyubov Laboratory of Theoretical Physics, Joint
Institute for Nuclear Research, Dubna  141980, Russia }
\author{M. Tater}
\affiliation{Institute of Nuclear Physics, Academy of Sciences of
the Czech Republic, \v{R}e\v{z} CZ--250 68, Czech Republic }
\author{E.~Truhl\'{\i}k}
\affiliation{Institute of Nuclear Physics, Academy of Sciences of
the Czech Republic, \v{R}e\v{z} CZ--250 68, Czech Republic }

\begin{abstract}
The capture of electrons by the nucleus $^{7}$Be from the
three-body initial state $p+e^{-}+^{7}{\rm Be}$ in the continuum
is studied. On the basis of the expansion of the three-body
continuum wave function in a small parameter
$\epsilon\approx\,\left({m_e}/{m_p}\right)^{1/2}$ [$m_e$ ($m_p$)
is the electron (proton) mass], the role of the protons on the
electron capture is considered. The results are compared with the
traditional treatment of the electron capture by the nucleus
$^{7}$Be. For stars with the density and temperature like in the
center of the Sun the studied mechanism can make non-negligible
contribution to the capture rate.
\end{abstract}

\noi 

\noi \hskip 1.9cm 

\maketitle

\section{Introduction}

The process of the electron capture by the nucleus $^{7}$Be is
important since it contributes to the low energy part of the
spectrum of neutrinos radiated by the Sun. Besides, it is obvious
that the balance of the disappearance channels of $^{7}$Be in the
Sun regulates the amount of the nucleus $^{8}$B which is the
source of the high energy solar neutrinos. This is the main reason
why this process attracted considerable attention over many years
\cite{b1,b2,b5,b6,b7,b9,b14}.

Practically all the discussion so far of the electron capture in
$^{7}$Be is concentrated on considering the electron wave function
in the vicinity of the nucleus and on the screening effects on it;
the study of this capture in the plasma was done in
Ref.\,\cite{b14}.

In what follows we will estimate the role of the process that is
usually not included in the standard theory of the $pp$ cycle in
the Sun. Let us first note that in the standard theory of this
cycle the destruction of the nucleus $^{7}$Be takes place in the
following binary reactions
\begin{eqnarray}
p\,+\,^{7}{\rm Be}\,&\rightarrow&\,^{8}{\rm B}\,+\,\gamma\,,  \label{br1} \\
e^-\,+\,^{7}{\rm Be}\,&\rightarrow&\,^{7}{\rm Li}\,+\,\nu\,.
\label{br2}
\end{eqnarray}
Since the nucleus $^{7}$Be participates in both processes, instead
of the binary reactions (\ref{br1}) and (\ref{br2}) we consider
the contribution to the electron capture rate from the
three-particle initial state \mbox{$p+e^{-}+^{7}$Be}. In this case
the following reactions can take place
\begin{eqnarray}
&\nearrow&\,^{7}{\rm Li}\,+\,\nu\,+p\,,    \label{3a} \\
p\,+\,e^{-}\,+\,^{7}{\rm Be}\,&\rightarrow&\,^{8}{\rm B}\,+\,\gamma\,+\,e^-\,, \label{3b}  \\
&\searrow&\,^{8}{\rm B}\,+\,e^-\,.  \label{3c}
\end{eqnarray}
As it was shown in Ref.\, \cite{b2} the screening corrections for
the electrons in the continuum are rather small. Therefore, we
consider in the initial state the bare Coulomb interaction in all
two-body subsystems $e^-+p$, $p+^{7}$Be, and $e^-+^{7}$Be. In this
case one can immediately realize that there is a qualitative
difference between the binary and ternary mechanisms\footnote{We
use the terms binary- and ternary reactions as synonyms for the
reactions in the two- and three component systems.} of the
electron capture. Indeed, if one starts from the three-body
initial state, then the processes (\ref{br1}) and (\ref{br2})
should be interdependent because the wave function of three
charged particles cannot be presented as a product of pair wave
functions, as is required by the binary processes (\ref{br1}) and
(\ref{br2})\footnote{There is only one exception corresponding to
the case when all three particles are at very large distances
between themselves \cite{b4}, which is not applicable to the
electron capture.}. Here we apply the method and results obtained
in Ref.\, \cite{b3} where an alternative to the Born-Oppenheimer
approach has been suggested for the calculations of wave functions
in the continuum for three charged particles.

\section{Calculations}

In analogy with the approach developed in Ref.\, \cite{b3}, the
continuum wave function of three charged particles can be expanded
in a small parameter $\epsilon$,
\begin{equation}
\epsilon\,=\,\left[\frac{M m_e}{(M+m_p)(m_p+m_e)}\right]^{1/2}
\,\approx\,\left[\frac{m_e}{m_p}\right]^{1/2}\,,  \label{eps}
\end{equation}
where, in addition to the electron and proton mass, also $M$, the
mass of the nucleus $^{7}$Be, enters. The expansion of the wave
function of the studied three-body system is then
\begin{equation}
\Psi(\vec r,\vec R)\,\approx\,\Psi_0(\vec r,\vec R)\,+\,
\epsilon\,\Psi_1(\vec r,\vec R)\,+\,\cdots\,.  \label{EXP}
\end{equation}
Since the parameter $\epsilon\,\approx\,0.0233$ one expects the
effects of the second term at the right hand side of
Eq.\,(\ref{EXP}) at the level of 2 \% in comparison with the first
term.

\begin{figure}[h!]
\centerline{ \epsfig{file=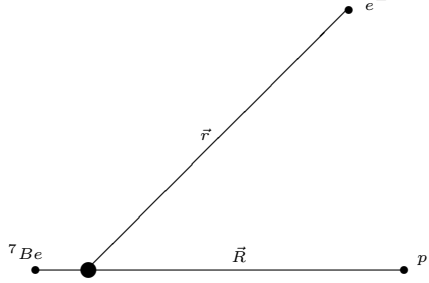} } \vspace{0.4cm} \caption{ The
Jacobi coordinates for the $p+e^{-}+^{7}$Be system.}\label{Fig1}
\end{figure}

It was found in \cite{b3} that in the limit
$\epsilon\,\rightarrow\,0$ the Jacobi coordinates $\vec r$ and
$\vec R$ (see Fig.\,\ref{Fig1}) separate. It means that the
structure of the wave function $\Psi_0(\vec r,\vec R)$ is
\begin{equation}
\Psi_0(\vec r,\vec R)\,=\,\Psi^C(\vec R)\,\Psi^C(\vec r,
Z=Z_1+Z_2)\,,  \label{Psi0}
\end{equation}
where $\Psi^C(\vec R)$ is the Coulomb wave function describing the
relative motion of the proton and the nucleus $^{7}$Be, and
$\Psi^C(\vec r, Z=Z_1+Z_2)$ is the Coulomb wave function that
describes the motion of the electron in the field of the effective
Coulomb potential of the charge $Z=Z_1+Z_2$.
The crucial point is that the wave function $\Psi^C(\vec r,
Z=Z_1+Z_2)$ depends on the distance between the electron and the
center of mass of the subsystem of heavy particles. It means that
even if the distance between the electron and the nucleus $^{7}$Be
is zero, as is required by the Hamiltonian of the weak
interaction, the wave function $\Psi^C(\vec r, Z=Z_1+Z_2)$,
defining the probability of the electron capture by the nucleus
$^{7}$Be, should be taken at a non-zero distance $|\vec r|=\beta
|\vec R|$, where $\beta=1/7$ is the ratio of the proton and
$^{7}$Be masses.

It is clear that this phenomenon appears due to the electron
movement in the Coulomb field of two charged particles with
positive charges. Following the arguments presented above we now
consider two effects acting in the opposite directions. On one
hand, increasing the effective positive charge of the heavy
particles system by one unit will enlarge the electron capture
rate. On the other hand, using the Coulomb wave function at finite
distances instead of the function taken at zero distance should
damp the capture rate.

Taking into account that the nuclear matrix elements of the
reactions (\ref{br2}) and (\ref{3a}) are the same, as a measure of
influence of the third particle (the proton in this case) on the
capture rate of electrons by the nuclei $^{7}$Be we introduce the
ratio ${\varsigma}(R, T)$, which is the function of the distance
$R$ between the particles and the temperature $T$,
\begin{equation}
{\varsigma}(R, T)\,=\,\frac{\int_0^\infty\,|\Psi^C_E(\beta R,
Z=5)|^2\,e^{-E/kT}\,dE}{\int_0^\infty\,F(Z=4,\nu)\,e^{-E/kT}\,dE}\,.
\label{Rb}
\end{equation}
Here the denominator on the right-hand side of the equation
contains the quantity that enters the electron capture rate from
the continuum for the reaction (\ref{br2}) \cite{b1}. The Fermi
function $F(Z,\nu)$ is given by the equation
\begin{equation}
F(Z,\nu)\,=\,2\pi\nu/(e^{2\pi\nu}-1)\,, \label{FF}
\end{equation}
where the parameter $\nu$ is given by the equation $\nu=-Z\alpha\,
m_e/p=-Z\alpha/v$, $\alpha$ is the fine structure constant, and
$p=\sqrt{2m_e E}$ is the electron momentum. It is obtained by
using the solution of the Dirac equation with the Coulomb
potential \cite{b15}.

Analogous integral in the numerator should reflect the effect of
the Coulomb potential on the electron in the continuum for the
reaction (\ref{3a}). For the wave function $\Psi^C_E(\vec r,Z)$,
we use the Coulomb continuum wave function for the state with zero
angular momentum
\begin{equation}
\Psi^C_E(\rho)\,=\,\frac{F_0(\eta,\rho)}{\rho}\,  \label{PCE}
\end{equation}
where the function $F_0(\eta,\rho)$ satisfies the equation
\begin{equation}
\frac{d^2 F_0}{d
\rho^2}\,+\,[\,1\,-\,\frac{2\eta}{\rho}\,]\,F_0\,=\,0\,,
\label{DEF0}
\end{equation}
with $\rho=p\beta R$, and
$\eta=-Z\alpha\sqrt{m_e/E}=-Z\alpha\sqrt{2}/v$ is the Sommerfeld
parameter. The function $F_0(\eta,\rho)$ can be expressed in terms
of the Kummer function $M$ (see Ref.\,\cite{b10}, Chap.\,14) as
\begin{equation}
F_0(\eta,\rho)\,=\,C_0(\eta)\,\rho\,e^{-i\rho}\,M(1-i\eta,2,2i\rho)\,,
\label{F0}
\end{equation}
where $C_0^2(\eta)=2\pi\eta/(e^{2\pi\eta}-1)$.

Instead of the quantity ${\varsigma}(R, T)$, one can consider
\begin{equation}
{\varsigma}_C(R, T)\,=\,\frac{\int_0^\infty\,|\Psi^C_E(\beta R,
Z=5)|^2\,e^{-E/kT}\,dE}{\int_0^\infty\,|\Psi^C_E(0,
Z=4)|^2\,e^{-E/kT}\,dE}\,.  \label{RCb}
\end{equation}
Using Eq.\,(\ref{DEF0}), one obtains
\begin{equation}
|\Psi^C_E(0, Z)|^2\,=\,2\pi\eta/(e^{2\pi\eta}-1)\,, \label{Psi0}
\end{equation}
which is of the same form as the Fermi function (\ref{FF});
however $\eta=\sqrt{2}\nu$.

One can find the following integral representation  (see
Ref.\,\cite{b10}, Chap.\,13)
\begin{equation}
M(1-i\eta,2,2i\rho)\,=\,\frac{sh \pi\eta}{\pi\eta}\,\int_0^1\,
e^{2i\rho t}\left(\frac{1-t}{t}\right)^{it}\,dt\,, \label{KF}
\end{equation}
for this function. In the numerical calculations of the integral
over energy in Eqs.\,(\ref{Rb}) and (\ref{RCb}), we used this
representation of the Kummer function for the energies $E>0.1$
keV. In the interval $E<0.1$ keV, the function under the integral
in Eq.(\ref{KF}) strongly oscillates which makes the calculations
difficult. Instead, we applied the program PFQ developed in Ref.\,
\cite{b11}. Let us note that for the energies $E>0.1$ keV the
program PFQ and Eq.\,(\ref{KF}) provide the same results to a high
degree of accuracy.

We also introduce the mean value $<{\varsigma}(R_0, T)>$ of the
function ${\varsigma}(R, T)$
\begin{equation}
<{\varsigma}(R_0, T)>\,=\,
N\int\,e^{-\frac{(R-R_0)^2}{2R^2_0}}\,{\varsigma}(R, T)\,d\vec R
\,=\,4\pi R^3_0 N \int_0^\infty e^{-\frac{(x-1)^2}{2}}x^2\,
{\varsigma}(R_0 x, T)\,dx \,, \label{BS}
\end{equation}
where
\begin{equation}
N^{-1}\,=\,\int\,e^{-\frac{(R-R_0)^2}{2R^2_0}}d\vec R
\,=\,\frac{4\pi R^3_0}{\sqrt{e}}[\,1\,+\,\sqrt{2\pi
e}\,(\,1\,-\,\phi(-\frac{1}{\sqrt{2}})\,)\,] \, \label{N}
\end{equation}
is the normalization constant. In Eq.\,(\ref{N}), the function
$\phi(y)$ is the error function (see Ref.\,\cite{b10}, Chap.\,7).
It can be seen from Eq.\,(\ref{BS}) that the quantity
$<{\varsigma}(R_0, T)>$ depends on the mean distance $R_0$ between
the particles defined by the density in the Sun and on the
temperature $T$. The mean value $<{\varsigma}_C(R_0, T)>$ is
defined analogously by using ${\varsigma}_C(R, T)$.

We checked the precision of numerical calculations  of the
quantities $<{\varsigma}(R_0, T)>$ and $<{\varsigma}_C(R_0, T)>$
by using Mathematica$^{\bigcirc\hspace{-2.4mm}{ R}}$ and also
independent numerical procedures. The results of these two
independent ways of calculations agree within the required
accuracy, which is \mbox{0.1 \%}.

\section{Results and discussion}

\begin{figure}[h!]
\centerline{ \epsfig{file=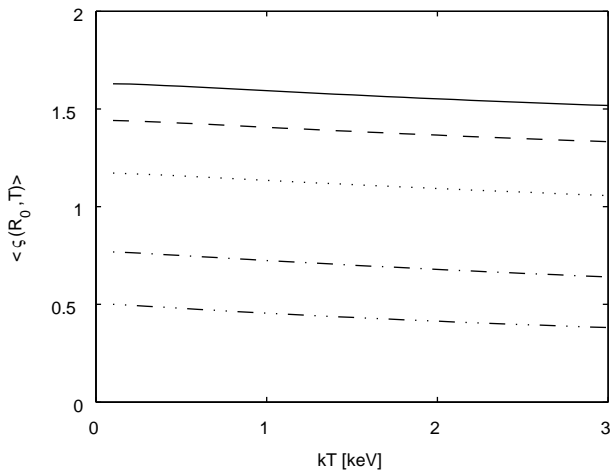} } \vspace{0.4cm} \caption{
The dependence of the mean value \mbox{$<{\varsigma}(R_0, T)>$} on
the temperature $T$ and the value of the mean  distance $R_0$.
Solid line, $R_0=0.1\times 10^4$ fm; dashed line, $R_0=0.25\times
10^4$ fm; dotted line, $R_0=0.5\times 10^4$ fm; dashed and dotted
line, $R_0=1.0\times 10^4$ fm; dashed and double dotted line,
$R_0=1.5\times 10^4$ fm.}\label{Fig2}
\end{figure}

\begin{figure}[h!]
\centerline{ \epsfig{file=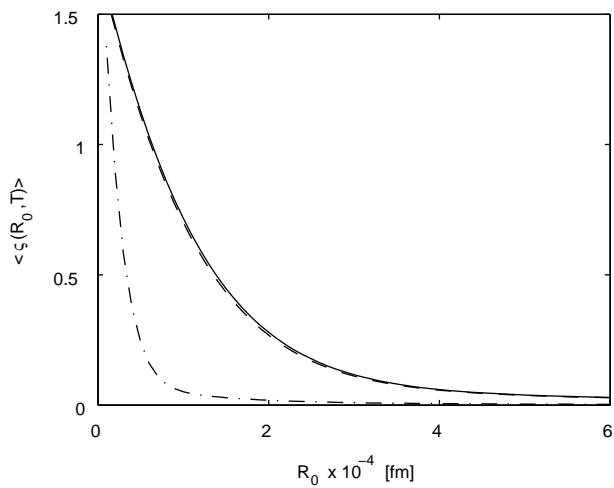} } \vspace{0.4cm} \caption{
The dependence of the mean value \mbox{$<{\varsigma}(R_0, T)>$} on
the values of the mean distance $R_0$. Solid curve, $kT=1.161$
keV, $\beta=1/7$, and $Z=5$; dashed and dotted curve, $kT=1.161$
keV, $\beta=4/7$, and $Z=6$; dashed curve, $kT=1.5$ keV,
$\beta=1/7$, and $Z=5$. The dashed and dotted curve corresponds to
analogous calculations for the nuclei $^{4}$He.}\label{Fig3}
\end{figure}

\begin{figure}[h!]
\centerline{ \epsfig{file=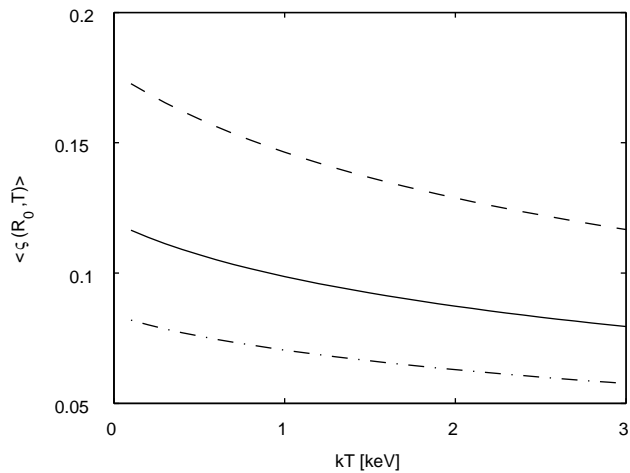} } \vspace{0.4cm} \caption{
The dependence of the mean value \mbox{$<{\varsigma}(R_0, T)>$} on
the temperature $T$ and the value of the mean  distance $R_0$.
Solid line, $R_0=3.25\times 10^4$ fm; dashed line, $R_0=2.75\times
10^4$ fm; dashed and dotted line, $R_0=3.75\times 10^4$
fm.}\label{Fig4}
\end{figure}

The results of the calculations are presented in Table I and
Figs.\,\ref{Fig2}-- \ref{Fig4}.
In Fig.\,\ref{Fig2}, the dependence of the mean value
\mbox{$<{\varsigma}(R_0, T)>$} given in Eq.\,(\ref{BS}) on the
temperature $T$ and the value of the mean distance $R_0$ is shown.
A weak dependence of \mbox{$<{\varsigma}(R_0, T)>$} on the
temperature means, as it follows from Eq.\,(\ref{Rb}), that the
temperature dependence of the electron capture by $^{7}$Be is
almost the same for the ternary and binary reactions. Such a
behaviour can be understood from the fact that in both cases only
rarely is all the kinetic energy carried by one particle, which is
the electron. On the other hand, the dependence of
\mbox{$<{\varsigma}(R_0, T)>$} on the value of $R_0$ shows that
the contribution to the capture rate of the ternary reaction is
presumably suppressed in stars  but it can be at the same level as
the contribution to the capture rate for the binary reaction or
even prevail over it at very high densities. This is natural
because at short distances between the particles the factor of the
larger effective charge acting on the electron will dominate. The
same conclusion can be drawn from Fig.\,\ref{Fig3}. In this
figure, the solid and dashed curves practically coincide. This
again shows a very smooth dependence of \mbox{$<{\varsigma}(R_0,
T)>$} on the temperature. Let us note that the values of the
$R_0$, considered in Fig.\,\ref{Fig2}, correspond to rather dense
stars. For example, the value of \mbox{$R_0=10^4$ fm} corresponds
to the proton density $\rho_p=1673$ g/cm$^3$, which is about 11
times larger than in the center of the Sun. Let us further discuss
the electron capture by $^{7}$Be solely in the Sun.

In Table I, we show the influence of protons on the electron
capture in the Sun in more detail. For the Standard Solar Model,
we choose Model SSMBP2004 \cite{b12}. According to Fig.\,6.1 of
Ref.\,\cite{b13}, the maximal intensity of the electron capture by
the nuclei $^{7}$Be takes place at the distance
$R_s/R_\odot\approx 0.06$, where $R_\odot$ is the radius of the
Sun, and it drops to one half at $R_s/R_\odot\approx 0.03$ and
$R_s/R_\odot\approx 0.1$. Using the data on the temperature, the
density, and the fraction of the hydrogen in this area of the Sun,
we obtain the mean values \mbox{$<{\varsigma}(R_0, T)>$} and
\mbox{$<{\varsigma}_C(R_0, T)>$} presented in Table I. In the
second column, we add the mean value calculated at $R_s/R_\odot$ =
0.007 which is close to the very centre of the Sun. It can be seen
that the change of the average quantities \mbox{$<{\varsigma}(R_0,
T)>$} and \mbox{$<{\varsigma}_C(R_0, T)>$} is very smooth.

Table I. The mean values \mbox{$<{\varsigma}(R_0, T)>$} and
\mbox{$<{\varsigma}_C(R_0, T)>$} for the electron capture by the
nuclei $^{7}$Be in the Sun.
\begin{center}
\begin{tabular}{|l||c|c|c|c|}\hline 
$<{\varsigma}(R_0, T)>$& 0.0991 & 0.0991 & 0.0965 & 0.0913 \\
$<{\varsigma}_C(R_0, T)>$& 0.0718 & 0.0717 & 0.0696 & 0.0658
\\\hline\hline
$R_s/R_\odot$& 0.007 & 0.03 & 0.06 & 0.1\\
kT (keV) & 1.353 & 1.300 & 1.161 & 1.088 \\
$\rho_p$ (g/cm$^{3}$)& 52.1 & 51.6 & 48.9 & 44.9 \\
 $R_0 \times 10^{-4}$ (fm)  & 3.179 & 3.188 & 3.250 & 3.340
\\\hline
\end{tabular}
\end{center}
As can be seen from the first row of Table I and from
Fig.\,\ref{Fig4}, the contribution to the capture rate of the
ternary reaction at the Sun is about 10 \% of the binary
one\footnote{The next term in the expansion of Eq.\,(\ref{EXP}) is
expected to change it only by $\approx$ 2 \%.}. This means that it
should increase sensibly the burning out of the nuclei $^{7}$Be in
comparison with the binary reaction, thus decreasing the
concentration of the nuclei $^{8}$B that appear after the capture
of protons by $^{7}$Be.

Comparing the first and the second rows of Table I shows a
difference of 3 \% between the values of \mbox{$<{\varsigma}(R_0,
T)>$} and \mbox{$<{\varsigma}_C(R_0, T)>$}. This variation arises
from the difference between the relativistic and nonrelativistic
estimations of the electron wave function at zero distance for the
binary reaction (\ref{br2}).

In Fig.\,\ref{Fig4} we show the variation of
\mbox{$<{\varsigma}(R_0, T)>$} for the reaction (\ref{3a}) for
larger intervals of $T$ and $R_0$.

One can consider in analogy the influence of the nuclei $^{4}$He
on the electron capture by $^{7}$Be in the Sun. In this case,
$Z=6$, $\beta=4/7$, and at the radius $R_s/R_\odot=0.06$ the mean
distance between the nuclei $^{4}$He is $R_0=5.34\times 10^4$ fm.
Then one obtains from Eq.\,(\ref{BS}) that
\mbox{$<{\varsigma}(R_0, T)>$}=0.0036, which is about 27 times
smaller that the analogous value of \mbox{$<{\varsigma}(R_0, T)>$}
for the protons given in the fourth column of Table I. Evidently,
this influence on the electron capture is negligible. This can
also be seen from Fig.\,\ref{Fig3}.

The main conclusion following from our calculations is that the
three-body process due to the presence of the proton in the
vicinity of the nucleus $^{7}$Be results in the capture of the
electron by an effective charge $Z=5$ instead of $Z=4$, which is
qualitatively the new effect that cannot be simulated by
introducing the Debye screening. This effect can increase the rate
of the electron capture from the continuum by $^{7}$Be  in the
Sun, which will reduce the concentration of the nuclei $^{8}$B
that appear after the capture of protons by $^{7}$Be.

\section*{Acknowledgments}
The discussion with J. Adam, Jr.\,is acknowledged. This work was
supported by grant GA CR 202/06/0746, by Project LC06002 of the
Ministry of Education, Youth and Sports of the Czech Republic, and
by the Votruba-Blokhintsev Program for Theoretical Physics of the
Committee for Cooperation of the Czech Republic with JINR, Dubna.

\end{document}